\documentclass[letter]{aa}

\usepackage{layouts}
\usepackage{booktabs}
\usepackage{graphicx}
\usepackage{microtype}
\usepackage{upgreek}
\usepackage{pgf}
\usepackage{float}
\usepackage{textcomp}
\usepackage{siunitx}
\usepackage{listings}
\usepackage{pythonhighlight}
\usepackage{array}
\usepackage{amsmath} 
\usepackage{multirow}
\usepackage{tabularx}
\usepackage{amssymb} 
\usepackage{tabularx}

\usepackage[colorlinks=true,linkcolor=blue,citecolor=blue, urlcolor=blue]{hyperref}

\usepackage{txfonts}
\newlength{\mytextsize}
\begin{document} 

   \title{PS$\,$J2107$-$1611: a new wide-separation, quadruply imaged lensed quasar with flux ratio anomalies}

   \subtitle{}

    \author{
    Frédéric~Dux\inst{1,3}, 
    Cameron~Lemon\inst{1}, 
    Frédéric~Courbin\inst{1},
    Dominique~Sluse\inst{2},
    Alain~Smette\inst{3},
    Timo Anguita\inst{4,5},
    Favio~Neira\inst{1,4}
    }
    \titlerunning{A quadruply lensed quasar with flux ratio anomalies}
    \authorrunning{Dux et al.}
    \institute{\email{frederic.dux@epfl.ch},\\
    Institute of Physics, Laboratoire d’Astrophysique, 
    École Polytechnique Fédérale de Lausanne (EPFL), 
    Observatoire de Sauverny, CH-1290 Versoix, Switzerland, 
    \and
    STAR Institute, University of Liège, Quartier Agora - Allée du six Août, 19c B-4000 Liège, Belgium 
    \and
    European Southern Observatory,  Alonso de Córdova 3107, Vitacura, Santiago, Chile 
    \and
    Instituto de Astrofisica, Facultad de Ciencias Exactas, Universidad Andres Bello, Av. Fernandez Concha 700, Las Condes, Santiago, Chile 
    \and 
    Millennium Institute of Astrophysics, Monseñor Nuncio Sotero Sanz 100, Oficina 104, 7500011 Providencia, Santiago, Chile 
    }

   \date{Received October 9, 2023; accepted October xx, 2023}

  \abstract
   {We report the discovery of PS$\,$J2107$-$1611,  a fold-configuration 4.3\arcsec-separation quadruply lensed quasar with a bright lensed arc. It was discovered using a convolutional neural network on Pan-STARRS \textit{gri} images of pre-selected quasar candidates with multiple nearby Pan-STARRS detections. Spectroscopic follow-up with EFOSC2 on the ESO 3.58m New Technology Telescope reveals the source to be a quasar at $z=2.673$, with the blended fold image pair showing deformed broad lines relative to the other images. The flux ratios measured from optical to near-infrared imaging in the Canada-France-Hawaii Telescope Legacy Survey, Pan-STARRS, the Legacy Surveys, and the Vista Hemisphere Survey are inconsistent with a smooth mass model as the fold pair images are $\sim$15 times too faint. Variability, time delay effects, and reddening are ruled out through multiple-epoch imaging and colour information. The system is marginally resolved in the radio in the Very Large Array Sky Survey S-band, where it has a 10 mJy detection. The radio flux ratios are compatible with the smooth mass macromodel. This system offers a unique tool for future studies of quasar structure with strong and microlensing. A more detailed analysis of follow-up with \textit{JWST}/MIRI, VLT/MUSE, VLT/ERIS, and data from the European Very Long Baseline Interferometer will be presented in an upcoming paper.
}
    
   \keywords{Gravitational lensing: strong,
             Galaxies: active
            }
   
   \maketitle
   
\section{Introduction}

Strongly lensed quasars, i.e.,  multiply imaged luminous active galactic nuclei (AGN) at cosmological distances, have been used extensively to address a broad range of astrophysical and cosmological phenomena. They are powerful probes of cosmological parameters, either in large statistical samples \citep[e.g.,][]{chae2003}, or individually through time-delay cosmography \citep[e.g.,][]{Refsdal1964, york2005, wong2020}. They constrain the lens galaxy mass distribution, from the largest spatial scales \citep[e.g.,][]{oguri2014} down to scale of dark matter substructure \citep[e.g.,][]{nierenberg2017}. They have been used to constrain the IMF of lens galaxies \citep[Initial Mass Function, e.g.,][]{schechter2014} through microlensing, the additional lensing effect on the quasar images by stars in the lensing galaxy. 
Lensed quasars are also powerful microlensing laboratories enabling to probe the quasar source structure:
the X-ray emitting region \citep[e.g.,][]{chartas2009}, the accretion disk size and temperature profile \citep[e.g.,][]{jimenez2014}, and the geometry of the broad-line region \citep[BLR, e.g.,][]{hutsemekers2019, hutsemekers2021, paic2022}. 
Finally, optical studies of the central kiloparsecs of quasar host galaxies at high-redshift are only possible with strongly lensed quasars as the host is not only amplified but also well-separated from the bright quasar images, into bright arcs \citep[e.g.,][]{bayliss2017}.

These science cases are currently hindered by the still limited number of confirmed lensed quasars, in particular quadruply imaged systems (quads), of which only $\sim$60 are known \citep{lemon2023}. 
A special sub-sample of quads are those that display radio emission; they can probe radio emission at high-redshift \citep[e.g.,][]{badole2022}, yield microlensing- and dust-free flux ratios, and VLBI (Very-Long-Baseline Interferometry) follow-up of such systems provide unprecedented datasets for constraining the lens galaxy mass \citep{spingola2018, powell2022}. 
Only $\mathcal{O}(10)$ radio-loud quads are known, mostly discovered over 20 years ago by the CLASS (Cosmic Lens All Sky Survey) and JVAS (Jodrell Bank VLA Astrometric Survey, \citealt{myers2003} and \citealt{Browne2003}). 
Since then, all quads discovered have been in the optical with only faint radio counterparts \citep[Table 5]{Hartley2021}.

Here, we report the discovery of an unusual new quad, discovered using a convolutional neural network (CNN) specifically designed to find well-separated quads in Pan-STARRS (PS) imaging data. This is the first discovery of a lensed quasar with a CNN. In this letter, we attempt to explain the unusual configuration of this quad through an analysis of available archival imaging data and new imaging and spectroscopy. 
We explain our method of discovery in Section~\ref{sec:discovery}, present the available data and follow-up spectroscopy in Section~\ref{sec:J2107}, and briefly draw conclusions from our analysis and state future plans in~\ref{sec:conclusions}.

Throughout this letter we assume a flat $\Lambda$CMD cosmology with $H_0=70{\rm \,km \,s^{-1}\, Mpc^{-1}}$ and $\Omega_m=0.3$.

\begin{figure*}[ht!]
    \centering
    \hspace{-1cm}
    \includegraphics{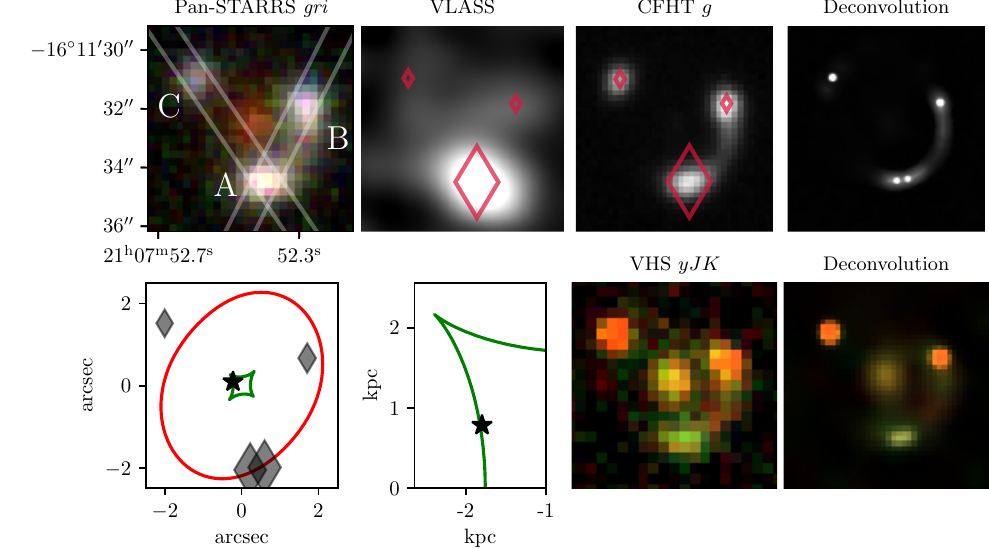}\vspace{-0.2cm}
    \caption{\textit{Top:} Pan-STARRS \textit{gri} image, with the lensed images labels and the two EFOSC2 slit positions used for the discovery spectra, followed by the VLASS radio map and the CFHT optical images. The last panel shows a deconvolution of the CFHT image with STARRED (see text).  The two central images are overlaid with markers of area proportional to the predicted magnifications from our smooth mass macro model of the lens.
    \textit{Bottom:} Mass model constrained using only the astrometry of the lensed images as derived with the CFHT data. A zoom on the point source position and the caustic in the source plane is shown on the second panel. The VHS \textit{yJK} image is shown on the third panel, followed by its STARRED deconvolution showing a noticeable dimming of the merging pair with wavelength in the near-IR compared to the optical.}
    \label{fig:imaging}
\end{figure*}

\section{Discovery method}
\label{sec:discovery}

The Pan-STARRS 3$\pi$ survey provides 1 arc-second seeing \textit{grizY} imaging of the entire sky North of a declination of $\delta=-$30$^\circ$ \citep{chambers2016}. This is an ideal dataset for finding bright gravitationally lensed quasars given its $r\sim$22  depth (5$\sigma$), which matches well the brightness of quasar samples from many surveys, such as \textit{Gaia} \citep[e.g.,][]{shu2019}.

Our initial catalogue is version 6.6\footnote{https://heasarc.gsfc.nasa.gov/w3browse/all/milliquas.html; from 14 June 2020} of MILLIQUAS \citep{flesch2021million}, which contains confirmed and candidate quasars compiled from the literature and recent quasar catalogues. 
To look for possible lensed systems, we cross-match to the Pan-STARRS catalogue, requiring 2 or more detections within 5\arcsec, resulting in $27\,760$ candidates, including 70 known lenses.

We train a CNN on mock quad lensed quasars to identify promising candidates. CNNs have been shown to be an effective method on recovering quads in mock datasets \citep{akhazanov2022}. Our mocks are built using the image and galaxy positions and $i$-band brightnesses provided in the catalogues of  \citet{oguri2010}. To evaluate the K-corrections in the $g$ and $r$ bands, we use the catalogue source and lens redshifts, and measure the corrections for the relevant filters using redshifted quasar and galaxy template from \citet{VandenBerk2001} and \citet{kinney1996} respectively. We do not include any reddening or microlensing effects. The light of the lens galaxy is represented as a Sersic profile~\citep{sersic}, and the quasar images as point sources. The image is then convolved with a Moffat~\citep{moffat1969} profile representing the Point Spread Function (PSF) with widths varying from 0.9 to 1.2 arcseconds. Finally, Poisson and Gaussian noise is added. For the negative examples, we used random systems with multiple detections in Pan-STARRS and with magnitudes matching those of our quasar sample.

The details of the training steps, CNN architecture, and application to updated quasar catalogues will be presented in an upcoming paper. 
We run the CNN on 10\arcsec\ Pan-STARRS \textit{gri} cutouts of the candidates centered on the MILLIQUAS catalogue position.
We visually inspect the cutouts of the 2000 best-scoring candidates, electing 131 promising ones. Among those, we recover 13 known quads, 23 doubles, and one new likely quad system PS$\,$J2107$-$1611. The first panel of Fig.~\ref{fig:imaging} shows the Pan-STARRS \textit{gri} image, 
The system displays 3 point-like components surrounding an extended object, with a clear arc visible between images A and B.
We estimate its total magnitude to $g=19.4$ and $i=18.4$. This is one of the brightest lensed arcs of all galaxy-scale lensed quasars.

\section{Follow-up and archival data}
\label{sec:J2107}

We acquired long-slit spectroscopy of PS$\,$J2107$-$1611 with EFOSC2 on the ESO 3.58m New Technology Telescope (NTT) in October 2020\footnote{P.I.: Timo Anguita, 106.218K.001}. Two exposures were taken with a 1.0\arcsec\ slit and grism \#13 ($R\sim260$), at position angles of PA=$-34.3^\circ$ and PA=$26.6^\circ$ East of North, as shown in the first panel of Fig.~\ref{fig:imaging}. The Differential Image Motion Monitor (DIMM) seeing was roughly 1.2\arcsec\ throughout the exposures. 

\begin{figure*}[ht!]
    \centering
    \input{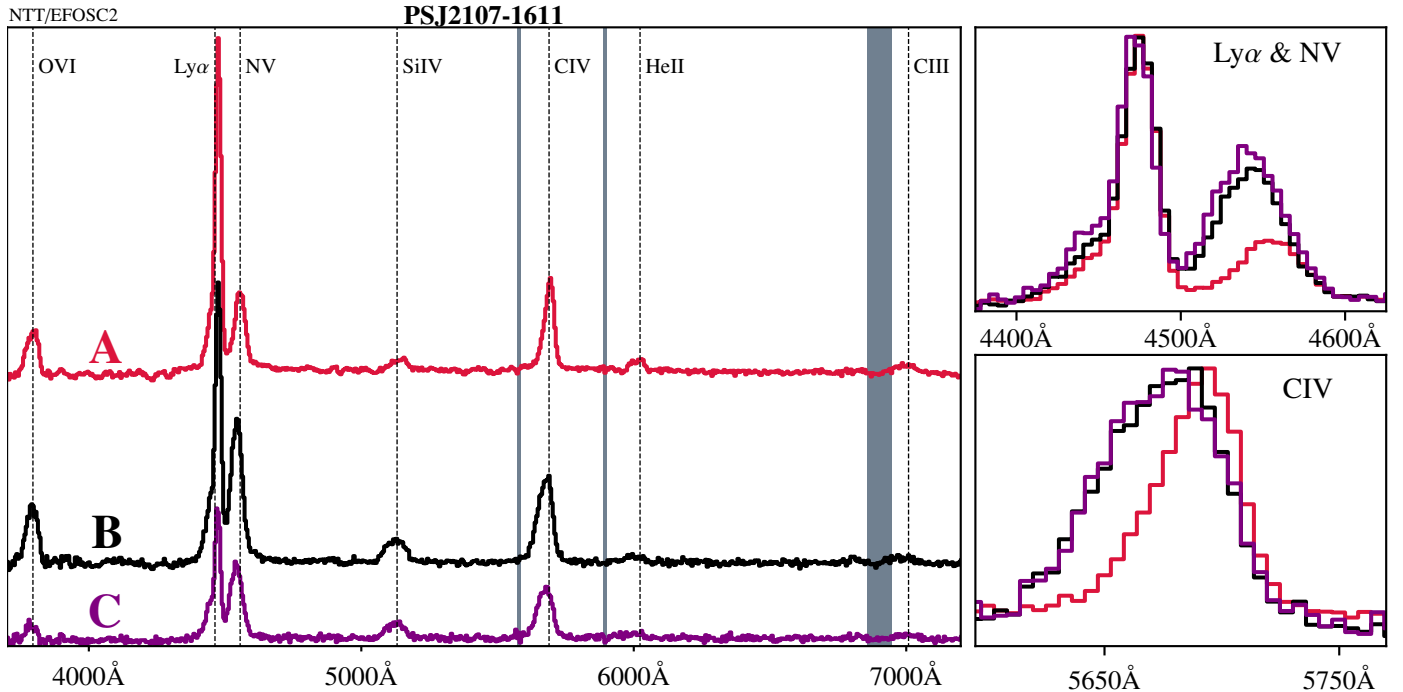}\vspace{-0.1cm}
    \caption{\textit{Left:} EFOSC2 spectra of the 3 visible lensed images of PS~J2107$-$1611. The A spectrum is the average of the individual spectra extracted from the two slit angles shown in Fig.~\ref{fig:imaging}.
    The fluxes were shifted for display purposes.
    \textit{Right:} Zoom-ins on the Ly$\alpha$, N\textsc{V} and C\textsc{IV} emission lines. The zoomed-in regions are normalized to their maximum intensity to highlight the larger skewness in these two lines of the A spectrum.}
    \label{fig:j2107spectra}
\end{figure*}

\begin{figure}
    \centering
    \includegraphics[]{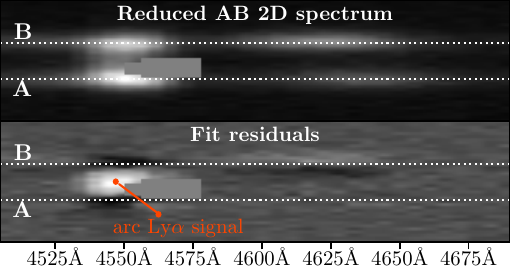}\vspace{-0.1cm}
    \caption{EFOSC2 2D spectrum for the slit orientation on images A and B. The Ly$\alpha$ emission line associated to the lensed quasar host galaxy is clearly detected between the A and B images. The arc is made visible after subtraction of the two point sources, on the bottom panel.}
    \label{fig:lyacontamination}
\end{figure}

The spectra are extracted as described in \cite{lemon2023}, providing modelled deblended spectra for two point sources along each slit. 
Since A is in both slit angles, we validate our extraction algorithm by checking that the independently extracted spectra of A match within the noise, and they do. 
The 3 spectra are typical of a quasar at $z_s \sim 2.673$ (Fig.~\ref{fig:j2107spectra}, where the 2 spectra of image A have been averaged). Images B and C are very similar with a flux ratio of $B/C\sim1.8$, approximately constant with wavelength and compatible with 2 images of the same object with differences attributable to extrinsic effects (e.g., microlensing).
There are however clear differences in the shapes of the \textsc{Nv} and \textsc{Civ} lines of A relative to B and C, as shown in the peak-normalised zoom-ins on the right panels of Fig.~\ref{fig:j2107spectra}. 
We also note the presence of clear residuals in the 2D A--B spectrum, between the two quasar images, at a wavelength corresponding to Ly$\alpha$ (Fig.~\ref{fig:lyacontamination}). We attribute this to the lensed host galaxy already seen in imaging.

To better understand the nature of this system, we analyse available archival imaging. In particular we find deep \textit{g}-band Canada-France-Hawaii~Telescope (CFHT) data with a total exposure time of 6901~seconds, an average seeing of 0.81\arcsec and a pixel size of 0.187\arcsec. We measure the astrometry of the different components using the 2-channel image deconvolution algorithm \textsc{STARRED}~\citep{starred}. \textsc{STARRED} models the bright point sources and a wavelet-regularized background as two channels and fits them jointly to the data. This idea has been shown to be very effective in doing precise photometry of blended point sources superposed on a complex background, i.e. the lensed quasars images, the lens galaxy, and the lensed quasar host galaxy in strongly lensed systems. Based on the MCS deconvolution method \citep{MCS}, \textsc{STARRED} is the method used to extract light curves of lensed quasars in the context of the COSMOGRAIL program ~\cite[see, e.g.,][]{Millon:2020ugy}.

Guided by preliminary lens models of the system and by the poorer quality of deconvolutions with 3 point-sources, we elect a model with 4 point-sources, which is also a general feature of all realistic lensing potentials \citep{burke1981}.  Image A is well compatible with being a blend of 2 images, A$_1$ and A$_2$, arising due to a source in fold configuration.
A single exposure of the CFHT dataset and the 4 point-source deconvolution is shown at the top-right of Fig.~\ref{fig:imaging}.
We take this as our fiducial deconvolved model as there is good evidence for it from both the lens model and the smooth resulting background arc. However, we note that higher resolution imaging is needed to confirm the fold pair. 

Next, we use the astrometry of the fiducial model given in Table~
\ref{tab:astrometry} to constrain a lensing mass model of a singular isothermal ellipsoid with external shear with \textsc{lenstronomy}~\citep{birrer2018lenstronomy, Birrer2021}. We do not use the flux ratios, and we fix the lens galaxy position to the centroid of the light as measured from the available $i$-band imaging data. The fit optimisation is done in the source plane with each image weighted equally. 
The associated total magnification is 78 with an Einstein radius of 2.12\arcsec. The ellipticity is 0.2 at 67 degrees East of North, and external shear strength of 0.06 at 115 degrees East of North. The individual magnifications are also plotted in the top-central panels of Fig.~\ref{fig:imaging}, illustrating the flux-discrepancy and compatibility in the optical and radio, respectively.

\begin{table}
\centering
\caption{Astrometry of the CFHT deconvolution, given for each point source relative to the lensing galaxy centered on RA(2000): 21h 07m 52.4s and DEC(2000):$-16^\circ$ 11\arcmin\ 32.4\arcsec (ICRS). We also include the magnification and time delays predicted by a mass model with a fiducial lens redshift of $z_l=0.5$ and constrained using this astrometry (see text). The point sources are labeled counter-clockwise.}\label{tab:astrometry}
\begin{tabular}{lrrrr}
\hline
& $\Delta\alpha \cos\delta $ [\arcsec] & $\Delta\delta$ [\arcsec] & Magnification & $\Delta t$ [days]\\
\hline
A$_1$ &  0.217  & -2.056 & $-33$ & 0 \\
A$_2$ &  0.593  & -1.993 & 36    & 0.05\\
B     &  1.703  &  0.668 & $-4$  & 20 \\
C     & -2.010  &  1.515 & 4     & -79 \\ 
\hline
\end{tabular}
\end{table}

We provide flux ratio estimates between the 3 visible point-sources (summing A$_1$ and A$_2$ together as A) measured from Pan-STARRS \textit{grizY}, Legacy Survey (LS) \textit{i}, and VHS \citep[VISTA Hemisphere Survey,][]{mcmahon2013} images. 
Due to the degeneracy between the flux originating from the ring and from the point sources, we measure the fluxes in the point sources with two methods: (i)~with a STARRED deconvolution, where a part of the total flux is absorbed by the wavelet-representation of the arc, and (ii)~with standard PSF photometry, where all the flux, including that of the arc, goes into the point sources. The two values provide a plausible range where the true flux values of the point sources should lie.
The filter, date, and B-A and C-A magnitude differences are listed in Table~\ref{tab:fluxratios}. The optical flux ratios are fairly constant across wavelength, followed by a decrease of the relative flux of A in the near-infrared bands. In Fig.~\ref{fig:imaging} we also show the VLASS image \citep[Very Large Array Sky Survey,][]{vlass}, which reveals that the brightest component is centred on A. We estimate a flux ratio of $\sim$15-to-1 between A and B, but this is highly uncertain given the large effective beam size and blending of the system.

\begin{table}
    \caption{Flux ratios measured on the optical images. 
    Two values are reported: the first stems from a STARRED deconvolution (PSF~+~wavelet background ring). The second, in parenthesis, stems from standard PSF photometry that sums the flux of the quasar images and the flux of the underlying arc (see text). 
    \label{tab:fluxratios}}
    \centering
    \begin{tabular}{lcrrr}
        \hline
        Band & Dataset & Epoch & \multicolumn{2}{c}{\(\Delta\mathrm{mag}\)} \\
        & & & \(B - A\) & \(C - A\) \\
        \hline
        g & CFHT & 2004-07-23  & -0.3 (0.4)   & -0.2  (1.0)  \\
        r & PS   & 2009-2014   & 0.4  (0.2)   & 0.5   (1.0)  \\
        i & PS   & 2009-2014   & 0.6  (0.4)   & 0.5   (1.2)  \\
        i & LS   & 2020-2021   & 0.5  (0.3)   & 0.6   (1.1)  \\
        Y & VHS  & 2014-04-15  & 0.4  (0.6)   & 0.3   (1.1)  \\
        J & VHS  & 2014-04-15  & -0.1 (0.5)   & -0.6  (0.6)  \\
        Ks & VHS & 2014-04-15  & -1.7 (-0.6) &  -2.1  (-0.9) \\
        \hline
    \end{tabular}
\end{table}

\section{Discussion and Conclusions}
\label{sec:conclusions}

Deformed broad lines as seen in the blended spectrum of image A relative to images B and C have been observed previously in other lensed quasars and attributed to microlensing \citep[e.g.][]{Richards2004, Sluse2011}. 
Studying the lines shapes in future spectra will be a good test of this, as changing shapes would be consistent with the variable nature of microlensing.
Microlensing is also a natural explanation of the highly discrepant measured flux ratios as compared with a smooth mass macromodel, 
though both the saddle point and minimum of the fold pair are 15 times too faint, which would be extremely unlikely \citep[see, e.g.,][]{weisenbach2021}. 
However, we must keep in mind that the image fluxes and positions from our deconvolution are prone to systematic uncertainty, given the limited seeing of the original data, and the image pair could indeed have very different fluxes. 
The marginally-resolved radio image of the system, however, does show flux ratios that agree with our best lens model, and should not suffer from microlensing given the much larger emission regions in the radio than in the optical/IR.

Several other effects were suggested to modify lensed images flux ratios: (i) dust reddening, but it is not a plausible explanation given the similar optical colours; (ii) a combination of variability and time delay effect, also an unlikely explanation given the lack of variability seen in individual Pan-STARRS epochs spanning several years (coupled with an estimated time delay of about 80 days); (iii) dark matter substructure \citep[e.g.,][]{McKean2007}; and (iv) finite source size effects coupled with differential magnification such as the radio source being well separated from the optical and infrared source \citep[see, e.g.,][]{barnacka2018, zhang2023}. 
It is likely that the latter two explanations possibly coupled with microlensing could explain the data, given that the source is constrained to be close to the caustic, potentially parsecs away, with a macro-magnification around 70. 

Further observations are required to fully understand PS$\,$J2107$-$1611 at high spatial resolution and with broad wavelength coverage and spectroscopy. 
Future work will present the analysis of such additional datasets of this system, including \textit{JWST}/MIRI, VLT/MUSE 2D spectra, VLT adaptive optics with the ERIS instrument, and in the radio with the European Very Long Baseline Interferometer (Lemon et al. in prep.).

\section{Acknowledgments}
This program is supported by the Swiss National Science Foundation (SNSF) and by the European Research Council (ERC) under the European Union’s Horizon 2020 research and innovation program (COSMICLENS: grant agreement No 787886).

TA acknowledges support from the Millennium Science Initiative ICN12\_009 and the ANID BASAL project FB210003.

\paragraph{Pan-STARRS:} The Pan-STARRS1 Surveys (PS1) and the PS1 public science archive have been made possible through contributions by the Institute for Astronomy, the University of Hawaii, the Pan-STARRS Project Office, the Max-Planck Society and its participating institutes, the Max Planck Institute for Astronomy, Heidelberg and the Max Planck Institute for Extraterrestrial Physics, Garching, The Johns Hopkins University, Durham University, the University of Edinburgh, the Queen's University Belfast, the Harvard-Smithsonian Center for Astrophysics, the Las Cumbres Observatory Global Telescope Network Incorporated, the National Central University of Taiwan, the Space Telescope Science Institute, the National Aeronautics and Space Administration under Grant No. NNX08AR22G issued through the Planetary Science Division of the NASA Science Mission Directorate, the National Science Foundation Grant No. AST-1238877, the University of Maryland, Eotvos Lorand University (ELTE), the Los Alamos National Laboratory, and the Gordon and Betty Moore Foundation.

\paragraph{CFHT:}
We made the use of observations obtained with MegaPrime/MegaCam, a joint project of CFHT and CEA/DAPNIA, at the CFHT which is operated by the National Research Council (NRC) of Canada, the Institut National des Science de l'Univers of the Centre National de la Recherche Scientifique (CNRS) of France, and the University of Hawaii. The observations at the Canada-France-Hawaii Telescope were performed with care and respect from the summit of Maunakea which is a significant cultural and historic site.

\paragraph{VLASS:}
The National Radio Astronomy Observatory is a facility of the National Science Foundation operated under cooperative agreement by Associated Universities, Inc. CIRADA is funded by a grant from the Canada Foundation for Innovation 2017 Innovation Fund (Project 35999), as well as by the Provinces of Ontario, British Columbia, Alberta, Manitoba and Quebec.

\paragraph{Legacy Surveys:}
The Legacy Surveys consist of three individual and complementary projects: the Dark Energy Camera Legacy Survey (DECaLS; Proposal ID \#2014B-0404; PIs: David Schlegel and Arjun Dey), the Beijing-Arizona Sky Survey (BASS; NOAO Prop. ID \#2015A-0801; PIs: Zhou Xu and Xiaohui Fan), and the Mayall z-band Legacy Survey (MzLS; Prop. ID \#2016A-0453; PI: Arjun Dey). DECaLS, BASS and MzLS together include data obtained, respectively, at the Blanco telescope, Cerro Tololo Inter-American Observatory, NSF’s NOIRLab; the Bok telescope, Steward Observatory, University of Arizona; and the Mayall telescope, Kitt Peak National Observatory, NOIRLab. Pipeline processing and analyses of the data were supported by NOIRLab and the Lawrence Berkeley National Laboratory (LBNL). The Legacy Surveys project is honored to be permitted to conduct astronomical research on Iolkam Du’ag (Kitt Peak), a mountain with particular significance to the Tohono O’odham Nation.

\bibliographystyle{aa}
\bibliography{refs}

\end{document}